\documentclass[a4paper,11pt]{article}
\usepackage{jinstpub} % for details on the use of the package, please see the JINST-author-manual
\usepackage{lineno}
\usepackage{siunitx}
\usepackage{svg}
\usepackage{cleveref}
\usepackage{subcaption}
\usepackage{lipsum}
\usepackage[nolist,nohyperlinks]{acronym}
\begin{acronym}
    \acro{LET}{linear energy transfer}
    \acro{CIRT}{carbon-ion radiotherapy}
    \acro{MKM}{microdosimetric kinetic model}
    \acro{SNR}{signal-to-noise ratio}
    \acro{CSA}{charge-sensitive amplifier}
    \acro{ESD}{electrostatic discharge}
    \acro{MOSFET}{metal–oxide–semiconductor field-effect transistor}
    \acro{ASIC}{application-specific integrated circuit}
    \acro{RBE}{relative biological effectiveness}
    \acro{ENC}{equivalent noise charge}
    \acro{MPV}{most probable value}
    \acro{TPS}{treatment planning system}
    \acro{NIRS}{national institute of radiological sciences (Japan)}
    \acro{LEM}{local effect model}
    \acro{PDK}{process design kit}
    \acro{JFET}{junction field-effect transistor}
\end{acronym}

\title{A frontend ASIC for Microdosimetry}

\author{Simon Waid$^1$, Matthias Knopf$^2$, Giulio Magrin$^3$, Albert Hirtl$^2$, Sebastian Onder$^1$, Stefan Gundacker$^1$, Daniel Radmanovac$^1$, Sandra Barna$^4$, Thomas Bergauer$^1$}

\affiliation{$^1$ Austrian Academy of Sciences, Marietta Blau Institute, Vienna, Austria}
\affiliation{$^2$ TU Wien, Institute for Atomic- and Sub-Atomic Physics, Vienna, Austria}
\affiliation{$^3$ MedAustron GmbH, Wiener Neustadt, Austria}
\affiliation{$^4$ Medical University of Vienna, Wien, Austria}

\emailAdd{simon.waid@oeaw.ac.at}

\abstract{Recent clinical evidence shows a correlation between \ac{LET} and tumor control in carbon ion radiotherapy. This prompts the direct inclusion of LET into the treatment planning. Currently, LET is mainly extracted from simulations. Good clinical practice requires adopting measurement routines that correlate with LET, such as microdosimetry. 
In this work, we describe an \ac{ASIC} for reading out microdosimeteric sensors. 
The \ac{ASIC} is designed for input capacitances up to \qty{3}{\pico\farad}. It contains four readout channels, each with a different saturation charge ranging from \qty{75}{\femto\coulomb} to \qty{3.2}{\pico\coulomb}. 
In the \qty{75}{\femto\coulomb} range, at \qty{1}{\pico\farad} input capacitance and a shaping time of \qty{1}{\micro\second}, the ASIC has an equivalent noise contribution (ENC) below \qty{15}{\elementarycharge} at ambient temperature. This low noise level is expected to enable new measurement possibilities,
including the assessment of microdosimetric proton spectra in the low–\ac{LET} region of the entrance channel, as well as studying the contribution of delta electrons.

}

\arxivnumber{2605.12008} % Only if you have one

\begin{document}
\maketitle
\flushbottom

\section{Introduction}
\acresetall
\label{sec:intro}
Recent clinical evidence in light ion radiotherapy shows that, besides the applied \ac{RBE} dose, the distribution of the \ac{LET} in the tumor impacts tumor control \cite{matsumotoUnresectableChondrosarcomasTreated2020,molinelliHowLEMbasedRBE2021,hagiwaraInfluenceDoseaveragedLinear2020}. %This prompts the inclusion of \ac{LET} into \ac{CIRT} treatment planning. 

This prompts the inclusion of LET and other radiation-quality parameters into \ac{CIRT} treatment planning. Such adoptions are already underway, either within commercial \acp{TPS}, e.g., \ac{LEM} I and \ac{MKM} \cite{bertoletImplementationMicrodosimetricKinetic2021} in RayStation, or through the development of an in-house clinical dose system, e.g., at \ac{NIRS} \cite{inaniwaReformulationClinicaldoseSystem2015}. Currently, \ac{LET} is extracted from simulations and exhibits significant inter-center variation \cite{hahnHarmonizingClinicalLinear2022}. Good clinical practice requires verification through measurements; microdosimetric detectors,
which can provide spectra of lineal energy—a quantity closely related to LET—are the
preferred detectors for this purpose.

Using microdosimetric sensors, lineal energy spectra are typically obtained from pulse-height analysis using charge-sensitive amplifiers, shapers, and multi-channel analyzers. Current microdosimetric equipment suffers from two main shortcomings: 
\begin{itemize}
    \item Noise: assuming a \qty{10}{\micro\meter} thick diamond microdosimeter, a 250 MeV proton in the entrance channel has an \ac{MPV} for the energy deposition of \qty{8.2}{\kilo\electronvolt}, %\footnote{\url{https://www.isotopea.com/webatima}} 
    corresponding to around \qty{625}{\elementarycharge} \cite{angelonePropertiesDiamondBasedNeutron2021}. A significant measurement would require an \ac{SNR} of at least 10 to ensure a clear distinction between signal and noise. Consequently, a readout frontend with an \ac{ENC} below \qty{62}{\elementarycharge} is required. Most frontends currently used in microdosimetry have a noise contribution well above \qty{100}{\elementarycharge}.
    \item Speed: Clinical adoption of microdosimetry requires measurements on unaltered clinical beams, with measurement times compatible with routine workflows, such as quality assurance. Thus, spectra need to be obtained within seconds to a few minutes, ideally simultaneously at multiple locations along the Bragg-Peak curve or at several locations within a phantom. 
    %Gib davor vielleicht noch einen Satz rein der die Challenge mit ein paar Beispielzahlen beschreibt: 
    % particle rates on the order of 1E9 on mm sized beamspots;
    % Pileup rejection capabilities on the order of 10 ns required;
    % While achieving good statistics of 1e5-1e6 cts (oder direkt hier den Parisi zitieren)
    This requirement is not fulfilled by any microdosimetric setup that the authors are aware of.
\end{itemize}

There is a speed-noise trade-off in charge-sensitive amplifiers \cite{bertuccioElectronicNoiseSemiconductorBased2023}, resulting in shaping times typically around \qty{1}{\micro\second} being chosen to balance noise and speed. The acquisition speed for low-pileup spectra is limited by this tradeoff. In addition, many amplifiers currently used in microdosimetry require a recovery time that is often longer than the shaping time to return to baseline. To address this limitation in acquisition speed, we propose implementing an array of microdosimeters operating in parallel. To further improve speed, we implemented an active reset circuit for baseline restoration, reducing recovery time.

To address the noise issue, we adopted the \ac{CSA} design methodology presented by Bertuccio and Caccia \cite{bertuccioNoiseMinimizationMOSFET2009}, further adopting design choices presented in \cite{bertuccioPossibilitySubelectronNoise2006,bertuccioProgressUltralownoiseASICs2007,bertuccioCMOSChargeSensitive2014, meleSIRIOHighSpeedCMOS2021,mansourFastLowJitterLowTimeWalk2023,degeronimoMOSFETOptimizationDeep2004,onderSiliconCarbideMonolithic2024}. 

\section{Circuit}

\begin{figure}
    \centering
    \includegraphics[width=0.75\linewidth]{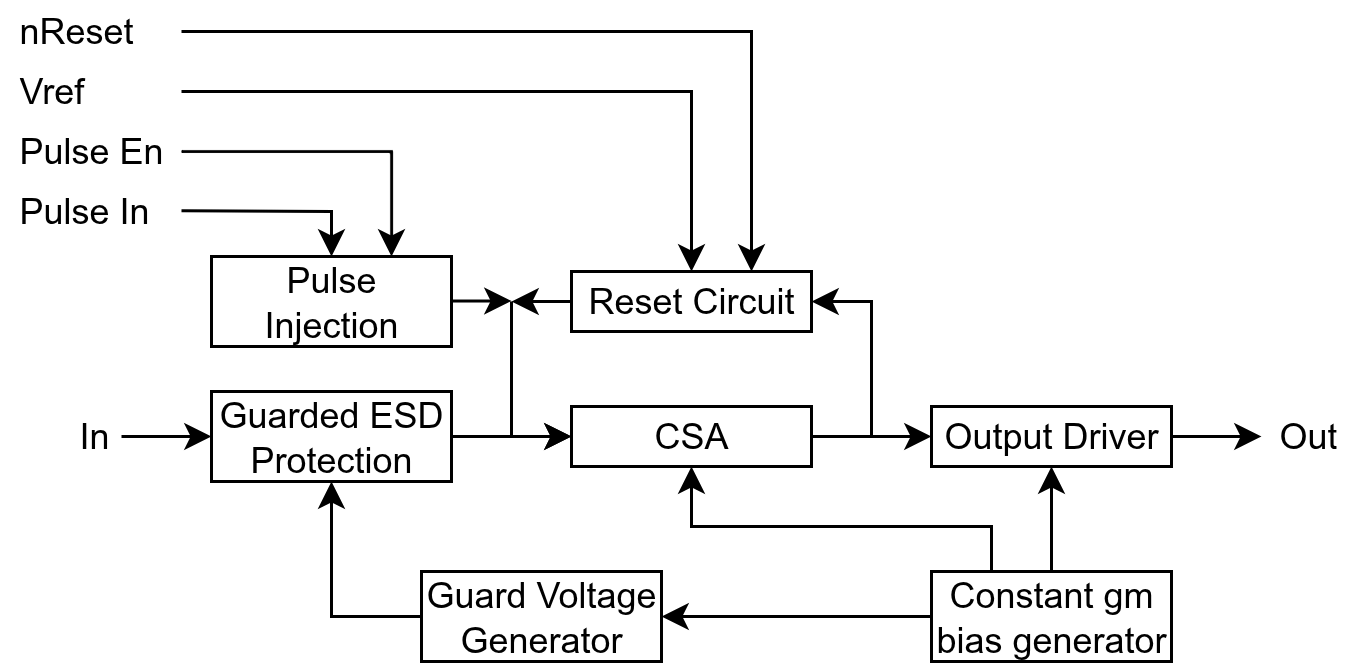}
    \caption{Frontend \ac{ASIC} block diagram.}
    \label{fig:block_diagram}
\end{figure}

The circuit block diagram for one readout channel of the microdosimetric frontend chip is shown in \Cref{fig:block_diagram}. The input is connected to a guarded \ac{ESD} protection circuit. It is supported by a guard-voltage generator that provides a voltage matching the \ac{CSA} input voltage. After the ESD protection circuit, the input reaches the CSA input. The CSA is supported by an active reset circuit, which restores the baseline of the \ac{CSA} output signal upon request. A switching circuit with charge-injection compensation detaches the reset circuit when not in use. This switching circuit also enables the attachment of an integrated charge-injection capacitor to inject test pulses. The \ac{CSA} output is buffered by an operational amplifier that drives the output pin. Biasing is provided by a simple constant-$g_\textrm{m}$ circuit to maintain circuit performance constant across operating temperatures. The constant-$g_\textrm{m}$ bias circuit increases the bias current with increasing temperature to compensate for the loss of transistor gain $g_\textrm{m}$. The consequence is reduced temperature sensitivity in the circuit's performance. The circuit was implemented with the open-source IHP SG130G2 \ac{PDK} and uses only thick-oxide \acp{MOSFET}. The operating voltage $V_\textrm{OP}$ is \qty{3.3}{\volt}.

\subsection{Charge Sensitive Amplifier}

\begin{figure}
    \centering
    \begin{minipage}[b]{0.53\textwidth}
        \centering
        \includegraphics[scale=0.83]{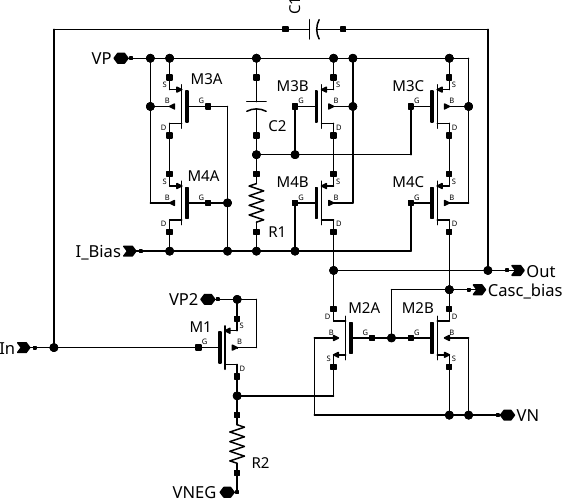}
        \caption{\ac{CSA} including biasing.}% \lipsum[1]}
        \label{fig:csa}
    \end{minipage}
    \hfill
    \begin{minipage}[b]{0.43\textwidth}
        \centering
        \includegraphics[scale=0.83]{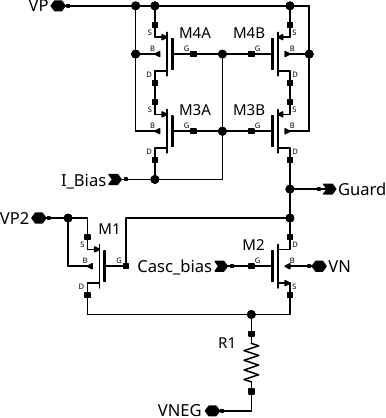}
        \caption{Guard voltage generator.}% \lipsum[1]}
        \label{fig:core_ref}
    \end{minipage}
\end{figure}

% \begin{figure}[h]
%     \centering
%     \begin{minipage}{0.47\textwidth}
%         \centering
%          \includegraphics[width=1\textwidth]{rise_time.pdf}
%         \caption{Output response to charge injection. The legend shows the size of the integration capacitor.}
%              \label{fig:rise_time}
%     \end{minipage}
%     \hfill
%     \begin{minipage}{0.47\textwidth}
%         \centering
%          \includegraphics[width=1\textwidth]{enc_vs_capa.pdf}
%         \caption{Simulated ENC vs input capacitance for selected shaping times at \qty{46}{\celsius}, typical process corner.}
%         \label{fig:enc_vs_capa}
%     \end{minipage}
% \end{figure}

The CSA, including parts of the biasing circuit, is shown in \Cref{fig:csa}. The input transistor was dimensioned following the procedure outlined in \cite{bertuccioNoiseMinimizationMOSFET2009}. The selected target input capacitance was 0.3-3 \unit{\pico\farad} with a nominal input capacitance of $C_\textrm{IL} = \qty{1}{\pico\farad}$. The gate capacitance $C_\textrm{G}$ was thus chosen to be $C_\textrm{G}=3 C_\textrm{IL}$. 
The oxide capacitance $C_\textrm{OX}'$, hole mobility $\mu_H$, and subthreshold slope $n$ were extracted from Spice simulations. The minimum length $L$ of a PMOS transistor in the selected process is \qty{0.4}{\micro\meter}. The determined values for a minimum length, thick oxide PMOS transistor are $C_\textrm{OX}'=\qty{4.98}{\femto\farad\per\square{\micro\meter}}$, $\mu_\textrm{H}=\qty{117}{\square{\centi\meter}\per{\volt\second}}$, and $n=1.46$. 
With $k$ as Boltzmann constant, $T$ as temperature and $q$ as elementary charge, the saturation current $I_\textrm{s}=2 n \mu C_\textrm{IL} (\frac{k T}{q L})^2$ at ambient temperature $T=\qty{300}{\kelvin}$ was determined to be $I_\textrm{S}=\qty{429}{\micro\ampere}$. With an inversion factor $R=5$, the biasing current was determined to be $I_\textrm{opt}=R I_\textrm{S}=\qty{2.15}{\milli\ampere}$. 

The cascode transistor of the \ac{CSA} (M2A in \Cref{fig:csa}) is biased by the current mirror consisting of M3A, M3B, M3C, and M4A, M4B, M4C. The current mirror was optimized for low noise, selecting a long-channel device with a length of \qty{10}{\micro\meter} for M3A, M3B, and M3C, and a minimum-length device for M4A, M4B, and M4C. Transistors M3A, M3B, M3C were operated at the maximum acceptable overdrive \cite{mansourFastLowJitterLowTimeWalk2023}, while the widths of M4A, M4B, M4C were a balance between speed and gain of the cascode and were thus optimized together with the cascode transistor M2A. The optimization goal was to transfer at least 90\% of the input charge to the integration capacitor C1, with an input capacitance of
\qty{3}{\pico\farad} at typical shaping times. This translates into a minimum voltage gain of \qty{62}{\decibel}. Further, the circuit was optimized for bandwidth, attaining the \qty{62}{\decibel} goal at around \qty{2}{\mega\hertz}. To enable optimal operation of the biasing circuit, the CSA's output swing was limited to \qty{2}{\volt}, leaving a window of \qty{1.3}{\volt} to the current sources. For the lower limit, \qty{0.5}{\volt} was chosen in order to keep M2A in the saturation region. Further, the negative supply voltage was chosen to enable the required voltage gain, resulting in $\textrm{VNEG} = \qty{-10}{\volt}$. Given the wide dynamic range needed in microdosimetry, CSAs with four different integration capacitors were implemented. The chosen capacitances are \qty{50}{\femto\farad}, \qty{200}{\femto\farad},  \qty{800}{\femto\farad}  and \qty{2.4}{\pico\farad}.

\subsection{ESD and Overvoltage Protection}

\begin{figure}
\centering
\begin{subfigure}{0.48\textwidth}
    \centering
    \includegraphics[scale=0.83]{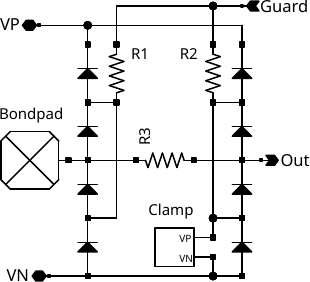}
    %\caption{ESD protection}
    %\label{fig:esd}
\end{subfigure}
\hfill
\begin{subfigure}{0.48\textwidth}
    \centering
    \includegraphics[scale=0.83]{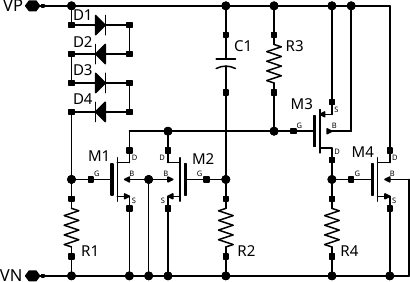}
    %\caption{Clamping  circuit}
    %\label{fig:clamp}
\end{subfigure}

\caption{Guarded ESD protection at the CSA input (left) and clamping circuit (right).}
\label{fig:esd_clamp}
\end{figure}

%{\color{red} Add component names to circuit}
% \begin{figure}
%     \centering
%     \includegraphics[width=0.5\linewidth]{core_ref_beautiful_bw.pdf}
%     \caption{Enter Caption}
%     \label{fig:core_ref}
% \end{figure}
The input \ac{ESD}/overvoltage protection circuit is shown in \Cref{fig:esd_clamp} (left). Two diodes in series clamp the input to the power rails before and after a resistor, as is commonly found in an ESD protection circuit for guarded inputs. However, in our case, circuit analysis revealed that this circuit might be insufficient for proper protection during operation. Assuming a current is flowing into the input, the input voltage may rise to twice the diode forward voltage above the operating voltage. Under this condition, the input transistor of the CSA (M1 in \Cref{fig:csa}) will turn off, and its source voltage will fall by one diode forward voltage below ground. 
Under this condition, the CSA input transistor will see the operating voltage plus three diode forward voltages between the gate and the source. At a nominal operating voltage of 3.3V +/- 10\%, this can exceed the guaranteed breakdown voltage of the gate oxide of 5.3 V. Consequently, a clamping circuit was introduced to limit the voltage at the gate of the input transistor. 

The clamping circuit is shown in \Cref{fig:esd_clamp} (right). If diodes D1 to D4 conduct, M1 will open, followed by M3, which provides additional amplification for opening M4. M4 will then clamp the guard voltage. For additional ESD protection, the capacitor C1 and the transistor M2 were added. They trigger the clamping circuit during ESD events.

The \Cref{fig:esd_clamp} (left) also requires a guard voltage matching the input voltage of the \ac{CSA}. To provide this guard voltage, a copy of the \ac{CSA}, shown in \Cref{fig:core_ref}, was implemented. For simplicity, the dimensions were set to match those of the main \ac{CSA}. For lower power consumption, a scaled-down version could also be used in the future.

\subsection{Reset}

\begin{figure}
    \centering
    \begin{minipage}[b]{0.6\textwidth}
        \centering
        \includegraphics[scale=0.83]{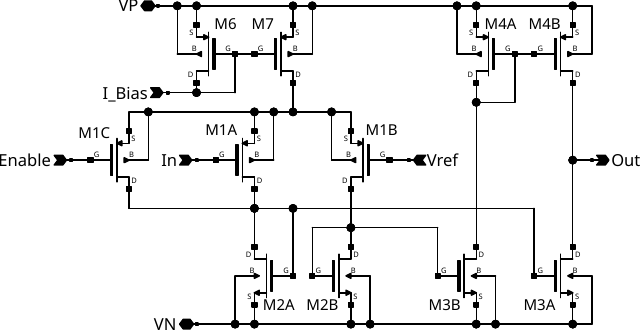}
        \caption{Active Reset circuit}
        \label{fig:reset_circuit}    
    \end{minipage}
    \hfill
    \begin{minipage}[b]{0.36\textwidth}
        \centering
        \includegraphics[scale=0.83]{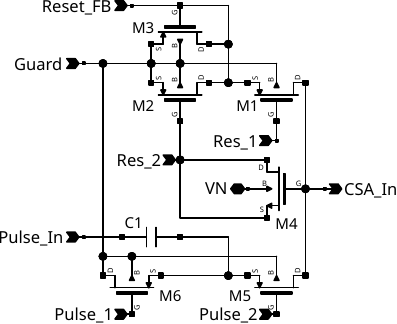}
        \caption{Switching circuit}
        \label{fig:switching}
    \end{minipage}
\end{figure}

An active reset circuit is employed to restore the \ac{CSA} baseline. The reset circuit is externally enabled and regulates the \ac{CSA}'s output voltage to match an externally provided reference. By varying the external reference voltage, the CSA can operate in both hole- and electron-collection modes. The reset circuit is shown in \Cref{fig:reset_circuit}. It consists of the difference amplifier M1A and M1B, which drive the rail-to-rail output driver consisting of M3A and M4B via current mirrors. Transistor M1C brings the circuit to a defined state when not in use, bypassing M1A and M1B, so that the biasing current, multiplied via the M3A-to-M2A width ratio, flows through M3A, while M4B remains closed. This prevents changes at the circuit's input from causing voltage transitions at its output, which could be capacitively coupled to the \ac{CSA} input via the switching network.

The output current of the reset circuit is matched to the integration capacitor to maintain a stable regulation loop while minimizing reset time. For this purpose, transistors M3A and M4B were scaled from 1 to 16 times their unitary size depending on the \ac{CSA} integration capacitor C1 in \Cref{fig:csa}. To simplify the layout, the circuit with 16 times unitary size was drawn first and used as a template for the other variants, removing transistors as needed.

% \begin{figure}
%     \centering
%     \includegraphics[scale=0.83]{pulse_reset_VarABC_beautiful.pdf}
%     \caption{Switching circuit}
%     \label{fig:switching}
% \end{figure}

When enabled, the reset circuit is attached to the CSA input via the switching network shown in \Cref{fig:switching}. This switching network also allows a charge-injection capacitor to be connected to the CSA input, enabling an external pulser to be connected to it. Both reset and pulse injection are performed in two phases, with a delay line controlling the timing. %The timing diagram is shown in fig. ??. 

To perform a reset operation, M2 in \Cref{fig:switching} is initially open, clamping the Reset\_FB node to the guard voltage. When a reset signal is received, M1 is first opened, connecting Reset\_FB to the \ac{CSA}'s input. Given that M2 is still open, this initially pulls the \ac{CSA} input to the guard potential, thereby bringing the voltage closer to the target value when the input voltage is far from it. Then, M2 is closed, and the reset circuit is enabled via the Enable signal in \Cref{fig:reset_circuit}. Now, the reset circuit's feedback loop acts on the CSA input, bringing the CSA output voltage to the desired value. During this time, M2 and M3 clamp the voltage at the Reset\_FB node to a reasonable range around the guard potential, ensuring that both the CSA and the reset circuit are at a reasonable operating point under all conditions. At the end of the reset operation, M1 is closed first, and M2 is opened again to prepare for the next cycle. Closing M1 injects a charge into the CSA, resulting in a baseline shift. This shift is compensated by injecting an opposite charge via M5 when Res\_2 transitions from high to low, opening M2 at the end of the cycle.

\begin{figure}
    \centering
    \includegraphics[scale=0.83]{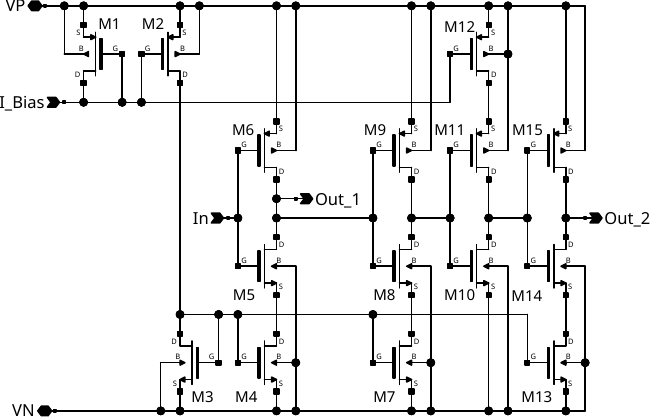}
    \caption{Delay Line}
    \label{fig:delay_line}
\end{figure}

To simplify the external interface, a delay line generating Res\_1 and Res\_2 from a single reset-enable signal was integrated on chip. The delay circuit shown in \Cref{fig:delay_line} and described in \cref{sec:delay_line}.

\subsection{Charge Injection}

The operation of the charge-injection circuit is similar to that of the reset circuit. When not in use, M6 (\Cref{fig:switching}) is open, bringing the node between M6, M5, and C1 to the Guard potential, which corresponds to the CSA input voltage. When the pulse injection circuit is enabled, M6 opens, freeing the node. Then, M5 is closed, attaching the node to the CSA input. Now, an external voltage step can be applied to the Pulse\_In input, causing charge injection into the CSA via capacitor C1. C1 was chosen to be identical to the integration capacitor in all cases.

Similar to the Res\_1 and Res\_2, the signals Pulse\_1 and Pulse\_2 controlling M5 and M6 are generated from a single input signal, using a delay line integrated on chip. The same delay line circuit as for the reset circuit is used and described in \cref{sec:delay_line}.

\subsection{Delay Line \label{sec:delay_line}}

In the delay line circuit, a transition from low to high (as needed to enable the reset circuit) transitions fast via M5, M9, M10, and M15 (\Cref{fig:delay_line}). The outputs Out\_1 and Out\_2 transition almost simultaneously, resulting in a smooth transition between clamping the reset circuit via M2 in \Cref{fig:reset_circuit} and controlling the CSA via the Reset\_FB node. Conversely, a transition from high to low to disable the reset circuit results in a fast transition from low to high at output Out\_1 via M6, while the transition from high to low at Out\_2 is delayed by limiting the current through transistors M8, M11, and M14 via M7, M12, and M13. Thus, it is ensured that after a reset, M1 in \Cref{fig:reset_circuit} is closed before M2 in \Cref{fig:reset_circuit} opens.

The circuit operates similarly for test-pulse injection. To enable the pulse injection, a transition from high to low at the input is required. This transition from high to low at the input results in M6 in \Cref{fig:reset_circuit} first closing before M5 in \Cref{fig:reset_circuit} opens. This ensures no current flows into the CSA via M6 in \Cref{fig:reset_circuit}. For the opposite direction, the timing is irrelevant, since after a pulse injection, the CSA is expected to be reset. Thus, Pulse\_1 and Pulse\_2 can transition simultaneously. 

To limit switching noise on the power rails from the inverters in the circuit, the cross-current is limited in all cases via current mirrors. This is desirable for M8, M11, and M14, as it results in a delay. In the case of M5, the resulting delay is neither needed nor harmful, as transitions from low to high at the inputs Out\_1 and Out\_2 switch contemporaneously.

% \begin{figure}
%     \centering
%     \includegraphics[width=0.5\linewidth]{pad_in_guard_beautiful_bw.pdf}
%     \caption{Enter Caption}
%     \label{fig:placeholder}
% \end{figure}
% \begin{figure}
%     \centering
%     \includegraphics[width=0.5\linewidth]{clamp_mini_beatiful_bw.pdf}
%     \caption{Enter Caption}
%     \label{fig:clamp}
% \end{figure}

%

\subsection{Output Driver}

\begin{figure}
    \centering
    \includegraphics[scale=0.83]{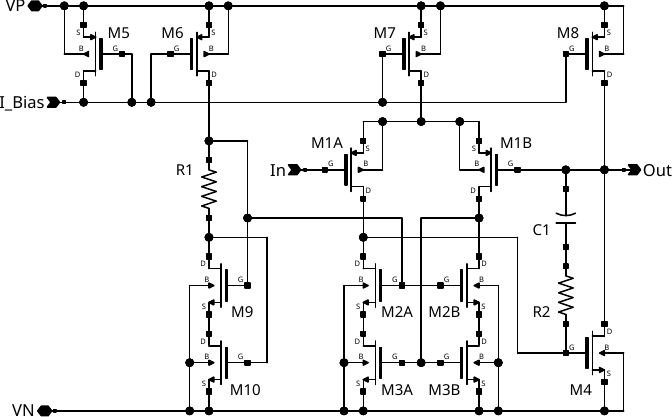}
    \caption{Output Driver}
    \label{fig:out_driver}
\end{figure}

The output driver is shown in \Cref{fig:out_driver}. It consists of a simple operational amplifier configured as a unity gain buffer. Together with the input pair M1A and M1B, the cascodes M2A and M3A, as well as  M2B and M3B, provide gain to drive the output transistor M4. C1 and R2 provide Miller compensation for unity-gain stability. The circuit is intended to be used with an additional external buffer and is thus designed to drive a load of only \qty{4}{\pico\farad}. The cross current of \qty{4}{\milli\ampere} in the output stage allows for resistive loads down to \qty{1}{\kilo\ohm}.

\section{Results and Discussion}

\begin{figure}[h]
    \centering
    \begin{minipage}[t]{0.48\textwidth}
        \centering
         \includegraphics[width=1\textwidth]{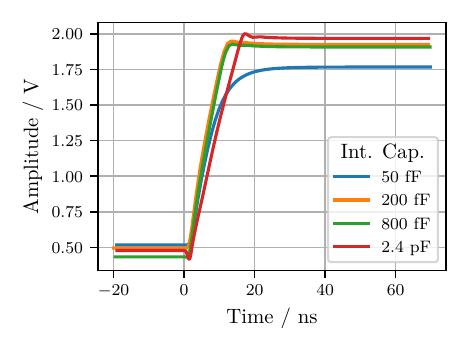}
        \caption{Output response to charge injection. The legend shows the size of the integration capacitor.}
             \label{fig:rise_time}
    \end{minipage}
    \hfill
    \begin{minipage}[t]{0.48\textwidth}
        \centering
         \includegraphics[width=1\textwidth]{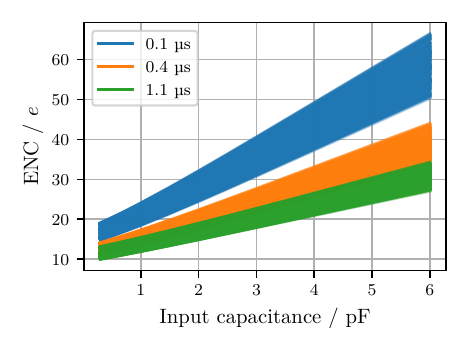}
        \caption{Simulated ENC vs input capacitance for an integration capacitor of \qty{50}{\femto\farad} for selected shaping times. Post layout simulation varying process corners, temperatures, and output voltages.}
        \label{fig:enc_vs_capa}
    \end{minipage}
\end{figure}

% \begin{figure}
%     \centering
%     \includegraphics[width=0.5\linewidth]{rise_time.pdf}
%     \caption{Output response to charge injection. The legend shows the size of the integration capacitor.}
%     \label{fig:rise_time}
% \end{figure}

% \begin{figure}
%     \centering
%     \includegraphics[width=0.5\linewidth]{enc_vs_capa.pdf}
%     \caption{Simulated ENC vs input capacitance for selected shaping times at \qty{46}{\celsius}, typical process corner.}
%     \label{fig:enc_vs_capa}
% \end{figure}

% \begin{figure}
%     \centering
%     \includegraphics[width=0.5\linewidth]{shaping_vs_output_voltage.pdf}
%     \caption{Minimum shaping time over output voltage and temperature}
%     \label{fig:shaping_vs_output_voltage}
% \end{figure}

\begin{figure}[h]
    \centering
    \begin{minipage}[t]{0.48\textwidth}
        \centering
        \includegraphics[width=1\textwidth]{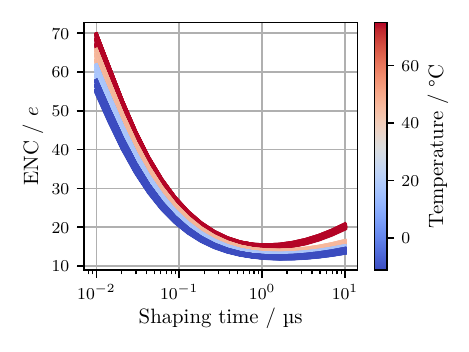}
        \caption{Simulated ENC as a function of shaping time and temperature for a CSA with \qty{50}{\femto\farad} integration capacitor, \qty{1}{\pico\farad} input load over process corners and output voltages.}
        \label{fig:enc_vs_shaping}
    \end{minipage}
    \hfill
    \begin{minipage}[t]{0.48\textwidth}
    
        \centering
        \includegraphics[width=1\textwidth]{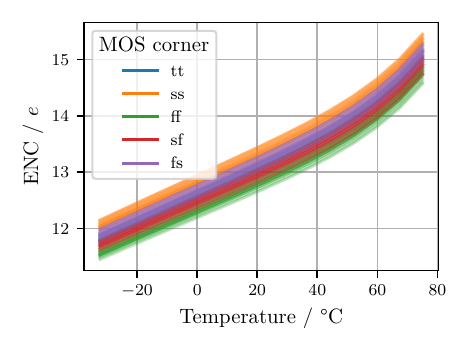}
        \caption{Simulated ENC as a function of temperature and process corner for a CSA with \qty{50}{\femto\farad} integration capacitor, \qty{1}{\pico\farad} input load,  \qty{1}{\micro\second} shaping for varying output voltages. ``MOS corner'' refers to the \ac{MOSFET} process corner.}
        \label{fig:enc_vs_temperature}
    \end{minipage}    
\end{figure}

% \begin{figure}
%     \centering
%     \includegraphics[width=0.5\linewidth]{shaping_vs_output_voltage.pdf}
%     \caption{Minimum shaping time over output voltage and temperature}
%     \label{fig:shaping_vs_output_voltage}
% \end{figure}

% \begin{figure}
%     \centering
%     \includegraphics[width=0.5\linewidth]{enc_vs_shaping.pdf}
%     \caption{Simulated ENC as a function of peaking time and temperature for a CSA with \qty{50}{\femto\farad} integration capacitor, \qty{1}{\pico\farad} input load over process corners and output voltages.}
%     \label{fig:enc_vs_shaping}
% \end{figure}

\begin{figure}[h]
    \centering
    \begin{subfigure}{0.48\textwidth}
        \centering
        \includegraphics[width=\textwidth]{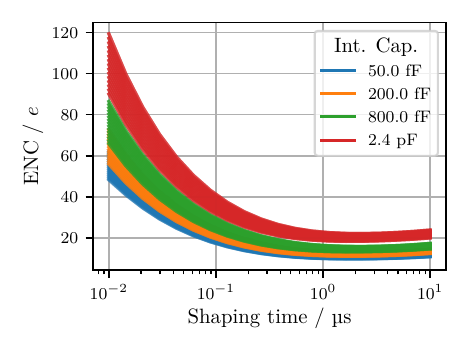}
        \caption{Pre-layout, after CSA}
        \label{fig:enc_all_capa_design_csa}
    \end{subfigure}    
    \hfill    
    \begin{subfigure}{0.48\textwidth}
        \centering
        \includegraphics[width=\textwidth]{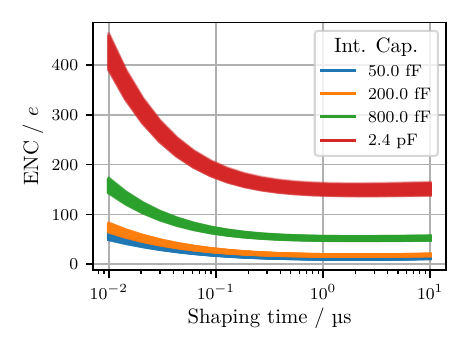}
        \caption{Pre-layout, at the output}
        \label{fig:enc_all_capa_design_out}
    \end{subfigure}
    \begin{subfigure}{0.48\textwidth}
        \centering
        \includegraphics[width=\textwidth]{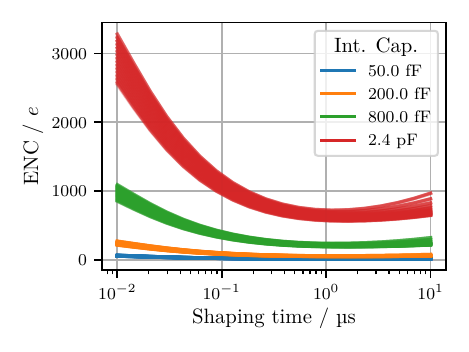}
        \caption{Post-layout, at the output}
        \label{fig:enc_all_capa_pex}
    \end{subfigure}   
    \caption{Simulated ENC pre- and post- layout before and after the output driver for \qty{1}{\pico\farad} input capacitance.  Simulations were carried out for different temperatures, process corners, and output voltages. Please note the different y-scales of the plots.} 
    \label{fig:enc_all_capa}

\end{figure}
%\hfill

% {\color{red}\begin{itemize}
%     \item ENC vs shaping 50 fF over temperatures,  ~ OK, enc\_vs\_shaping
%     \item Max shaping vs output voltage, 50 fF ~ Needs corners    
%     \item ENC over input capacitance, 50 fF OK enc\_vs\_capa  
%     \item ENC over temperature, 50 fF, 1µs, OK enc\_over\_temperature
%     \item ENC vs shaping, all capa, before output amplifier.
%     \item ENC vs shaping, all capa, after output amplifier. 
%     \end{itemize}
%     }

%\unit{\micro\second}

At the time of writing, the physical chip was still in production, so no measurements could be performed. Post layout simulations are thus used instead. They reveal that in the case of a \qty{50}{\femto\farad} integration capacitor at ambient temperature, and with an input capacitance of \qty{1}{\pico\farad}, an \ac{ENC} of \qty{15}{\elementarycharge} can be reached at a shaping time of \qty{1}{\micro\second}. The dependence of the \ac{ENC} for the  \ac{CSA} with \qty{50}{\femto\farad} integration capacitor on shaping time, temperature, and input capacitance is shown in \Cref{fig:enc_vs_capa}, \Cref{fig:enc_vs_shaping}, and \Cref{fig:enc_vs_temperature}. 

For larger integration capacitances, the \ac{ENC}, as a function of shaping time, is shown in \Cref{fig:enc_all_capa}. \Cref{fig:enc_all_capa_design_csa} shows the \ac{ENC} for pre-layout simulations directly after the \ac{CSA}, while \Cref{fig:enc_all_capa_design_out} shows the \ac{ENC} for pre-layout simulations at the output of the chip. Comparing the two figures (please note the different y-axis scaling), one can see that immediately after the \ac{CSA}, the \ac{ENC} is very similar across different integration capacitors, whereas at the output, there are substantial differences. Thus, for larger integration capacitances, the noise is dominated by the output driver and not the \ac{CSA}. The situation at the output further deteriorates when parasitics are considered, as shown in \Cref{fig:enc_all_capa_pex}. While this behavior is expected and acceptable, the high dynamic range observed at the \ac{CSA} could open a possibility of increasing the dynamic range in future revisions. Instead of using \ac{CSA}s with different integration capacitances, one could implement an amplifier with adaptive gain after the \ac{CSA}, thereby selecting the amplifier's gain for each event. Due to the strict timeline that needed to be followed in the current design, this could not be implemented yet.

\begin{figure}
    \begin{minipage}[t]{0.48\textwidth}  
        \centering
        \includegraphics[width=1\linewidth]{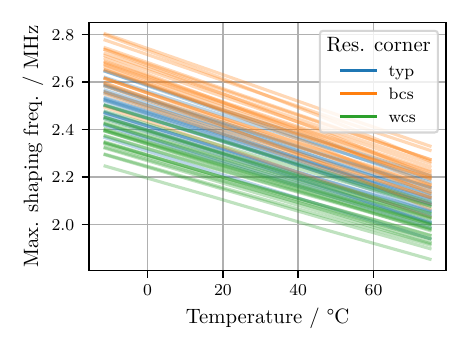}
        \caption{Maximum frequency at which 90\% of charge is collected by the \qty{50}{\femto\farad} integration capacitor as a function of the temperature. Simulations were carried out for different process corners and output voltages. ``Res. corner'' refers to the resistor process corner.}
        \label{fig:ac_fmax_shaping}
    \end{minipage}
     \hfill    
    \begin{minipage}[t]{0.48\textwidth}  
        \centering
        \includegraphics[scale=0.83]{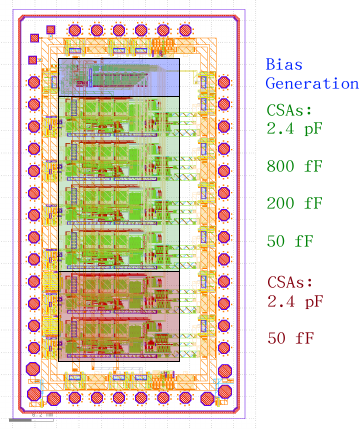}
        \caption{Floor plan of the die. Areas are delineated by colorful boxes. At the top, the bias generation circuit is visible (blue), followed by four \acp{CSA} (green) with \qty{2.4}{\pico\farad}, \qty{800}{\femto\farad}, 200 \qty{200}{\femto\farad}, and \qty{50}{\femto\farad} integration capacitor. Finally, two additional \acp{CSA} (red) with \qty{2.4}{\pico\farad} and \qty{50}{\femto\farad} integration capacitors were added to fill the available space.}
        \label{fig:die}
    \end{minipage}

\end{figure}

The usable bandwidth when requiring  90\% of charge being collected by the integration capacitor for the \qty{50}{\femto\farad} integration capacitor is shown in \Cref{fig:ac_fmax_shaping}. It is at around \qty{2}{\mega\hertz} for the \qty{50}{\femto\farad} integration capacitor, corresponding to a shaping time of around \qty{250}{\nano\second}. This is the recommended minimum shaping time for amplitude measurements. Lower shaping times are possible, however, at the cost of lower amplitude stability and increased noise. For larger integration capacitors, the minimum shaping time decreases.

The minimum shaping time for pileup rejection can be assessed from transient simulations. \Cref{fig:rise_time} shows a transient simulation of a charge-injection pulse corresponding to the feedback capacitance times \qty{1.5}{\volt}, with the input loaded with \qty{3}{\pico\farad} and the output loaded with \qty{4}{\pico\farad} for all types of integration capacitors. Except for the \qty{2.4}{\pico\farad} integration capacitor, 80 \% of the final value is reached after \qty{10}{\nano\second}. After \qty{20}{\nano\second} we are already within $\pm$ 5\% of the final value.  Thus, it is reasonable to implement pileup rejection with shapers with a shaping time down to \qty{10}{\nano\second}, while maintaining shaping times for peak detection in the range from \qty{250}{\nano\second} to \qty{1}{\micro\second}.

The floor plan of the \ac{ASIC} is shown in \Cref{fig:die}. Overall, the die contains six CSAs, with the version at \qty{2.4}{\pico\farad} and \qty{50}{\femto\farad} placed twice to fill the available space. Each CSA has its own VNEG pin (\cref{fig:csa}), while VP and VP2 are shared. Disconnecting the VNEG pin of unused CSAs can reduce power consumption. If all six CSAs have their VNEG pins connected, the ASIC is expected to consume up to \qty{500}{\milli\watt}. With only one CSA operational, the consumption can be reduced to around \qty{150}{\milli\watt}.

\begin{table}
    \centering
    \begin{tabular}{ccccccc}
         Source &  Input Cap. & Shaping Time &  ENC & Process &Notes  \\
         This work  &  \qty{0.3}{\pico\farad} to  \qty{3}{\pico\farad}  & \qty{0.25}{\micro\second} to \qty{2}{\micro\second} &  \qty{15}{\elementarycharge}  to \qty{40}{\elementarycharge}  & Equiv.  \qty{400}{\nano\meter} & \\
         \cite{drobizhevCharacterizationFirstPrototype2026} & \qty{3}{\pico\farad}  & \qty{0.1}{\micro\second} to \qty{100}{\micro\second}
         & \qty{92}{\elementarycharge} to \qty{150}{\elementarycharge} &  \qty{65}{\nano\meter} & At \qty{300}{\kelvin}  \\
         \cite{haoLowNoiseWide2024} & \qty{2}{\pico\farad}  & \qty{1}{\micro\second} to \qty{20}{\micro\second} &  Measured: \qty{42}{\elementarycharge}  &  \qty{180}{\nano\meter}  & \\
         \cite{buttaDesignCharacterizationLowNoise2024} & \qty{2}{\pico\farad}  &  \qty{3}{\micro\second}  & \qty{32}{\elementarycharge} &  Not given & \\
         \cite{trigilioETTORE12ChannelFrontEnd2018b} & \qty{3}{\pico\farad}  &  Down to \qty{0.03}{\micro\second}   & Down to \qty{5.3}{\elementarycharge} & 
         \qty{350}{\nano\meter}, JFET & At \qty{-30}{\celsius}\\
    \end{tabular}
    \caption{Comparison of our design to \acp{CSA} designs found in the literature.}
    \label{tab:comparison}
\end{table}

\Cref{tab:comparison} compares this work to other \acp{CSA}. Our design has a similar performance to comparable \ac{MOSFET}-based solutions. In the comparison, one design stands out, exhibiting substantially better performance, namely the design by Trigilio et al. \cite{trigilioETTORE12ChannelFrontEnd2018b}. This can be attributed to the use of a \ac{JFET} in the input stage.

\section{Conclusion}
In this work, we presented a front-end \ac{ASIC} for reading out microdosimeters, designed to address the main shortcomings of state-of-the-art microdosimetry readout circuits: speed and noise. The design is based on the  \ac{CSA} design methodology proposed by Bertuccio and Caccia \cite{bertuccioNoiseMinimizationMOSFET2009}. The design is expected to enable an \ac{ENC} at a shaping time of \qty{1}{\micro\second} and an input capacitance of \qty{1}{\pico\farad} below \qty{15}{\elementarycharge}, while still maintaining an \ac{ENC} below \qty{30}{\elementarycharge} at \qty{1}{\pico\farad} and staying below \qty{60}{\elementarycharge} for  shaping times down to \qty{250}{\nano\second}. The ASIC is expected to enable new measurement capabilities, such as measuring proton spectra at the entrance channel and studying delta-electron contributions in lineal energy spectra.

The designed \ac{ASIC} contains \acp{CSA} with four different feedback capacitors, whereby only the \ac{CSA} with the smallest feedback capacitor of \qty{50}{\femto\farad} attains the above-described performance, making only this variant suitable for studying delta-electrons and protons on the entrance channel. The \acp{CSA} with larger feedback capacitors exhibit a larger noise, but provide sufficient dynamic range to also study high-energy events such as the fragmentation tail of carbon beams. Their noise is limited by the output stage, a limitation that could be addressed in future versions of the \ac{ASIC} by introducing a variable-gain output stage to increase dynamic range. Finally, in most applications, the \ac{ASIC} will require an output buffer nearby, as the output stage has a limited driving capability of capacitive loads up to \qty{4}{\pico\farad}.

\section{Acknowledgment}
This project has received funding from the Austrian Research Promotion Agency FFG, Austria, grant number 918092.

\bibliography{HEPHY-detector-dev}
\bibliographystyle{ieeetr}

\end{document}